\documentclass[12pt]{article}
 \usepackage{epsfig}
\usepackage[latin1]{inputenc}

\usepackage[english]{babel}

\def\lsim{\,\raise0.3ex\hbox{$<$\kern-0.75em\raise-1.1ex\hbox{$\sim$}}\,}
\def\gsim{\,\raise0.3ex\hbox{$>$\kern-0.75em\raise-1.1ex\hbox{$\sim$}}\,}

\newcommand{\nuCC}{\nu, {\rm CC}}
\newcommand{\nubarCC}{\bar \nu, {\rm CC}}
\newcommand{\nunubarCC}{\nu(\bar \nu), {\rm CC}}
\newcommand{\nuNC}{\nu, {\rm NC}}
\newcommand{\nubarNC}{\bar \nu, {\rm NC}}
\newcommand{\nunubarNC}{\nu(\bar \nu), {\rm NC}}
\newcommand{\bea}{\begin{eqnarray}}
\newcommand{\eea}{\end{eqnarray}}
\newcommand{\beq}{\begin{equation}}
\newcommand{\eeq}{\end{equation}}

\begin{document}

\begin{titlepage}
\begin{flushright}
21 March, 2006\\
HIP-2006-15/TH\\
hep-ph/0603155
\end{flushright}
\begin{centering}
\vfill

{\large\bf NuTeV $\sin ^2 \theta _{\rm W}$ anomaly and nuclear parton distributions revisited}

\vspace{0.5cm}
 K. J. Eskola\footnote{kari.eskola@phys.jyu.fi} and 
 H. Paukkunen\footnote{hannu.paukkunen@phys.jyu.fi} \\

\vspace{1cm}
{\em Department of Physics,
P.B. 35, FIN-40014 University of Jyv\"askyl\"a, Finland}\\
\vspace{0.3cm}
{\em Helsinki Institute of Physics,
P.B. 64, FIN-00014 University of Helsinki, Finland}\\

\vspace{1cm}\centerline{\bf Abstract}
\end{centering}
\vspace{0.3cm}

By studying the Paschos-Wolfenstein (PW) ratio of deep inelastic 
$\nu{\rm Fe}$ and $\bar \nu{\rm Fe}$ scattering cross sections, we show that 
it should be possible to explain the NuTeV $\sin ^2 \theta _{\rm W}$ anomaly with quite 
conventional physics, by introducing mutually different nuclear modifications
for the valence-$u$ and valence-$d$ quark distributions of the protons 
in iron. Keeping the EKS98 nuclear modifications for $u_V+d_V$ as a baseline,
we find that some 20-30 \% nuclear modifications to the $u_V$ and $d_V$ distributions 
account for the change induced in the PW ratio by the NuTeV-suggested increase
$\Delta \sin ^2 \theta _{\rm W}=0.005$. We show that introduction of such
nuclear modifications in $u_V$ and $d_V$ individually, does not lead into contradiction with the 
present global DGLAP analyses of the nuclear parton distributions, where deep inelastic $lA$ 
scattering data and Drell-Yan dilepton data from $pA$ collisions are used as constraints. 
We thus suggest that the NuTeV result serves as an important further constraint in 
pinning down the nuclear effects of the bound nucleon PDFs. We also predict that if the 
NuTeV anomaly is explained by this mechanism, the NOMAD experiment should see an increase
in the weak mixing angle quite close to the NuTeV result.

\vfill
\end{titlepage}

%%%%%%%%%%%%%%%%%%%%%%%%%%%%%%%
%%%%  INTRODUCTION
%%%%%%%%%%%%%%%%%%%%%%%%%%%%%%%
\section{Introduction}

A few years ago, the NuTeV collaboration at Fermilab reported that their measurements 
in deep inelastic neutrino-iron scattering indicate the value of Weinberg weak mixing 
angle $\sin^2\theta_W = 0.2277 \pm 0.0013 ({\rm stat}) \pm 0.0009 ({\rm syst})$ 
\cite{Zeller:2001hh}. This result was surprising, being about $3\sigma$ above the world average 
$\sin^2\theta_W = 0.2227 \pm 0.00037$ \cite{Abbaneo:2001ix}. 
A number of possible solutions to this deviation --- 
the 'NuTeV anomaly' --- has been proposed, ranging from conventional 
explanations within the Standard Model (SM) to more exotic ones requiring 
novel, beyond-SM, physics. For reviews, see Refs.~\cite{McFarland:2003jw,Davidson:2001ji}.
Today, the NuTeV anomaly is still an open problem calling for an answer.

As intriguing it would be to see new physics appearing, it is obviously crucial to first 
investigate in detail whether the explanation could be hidden in the
features of parton distribution functions (PDFs) of the free and bound nucleons
which are not yet fully known. 
For instance, the possible local asymmetry between strange and antistrange quark 
distributions, $S^- \equiv \int dx x[s(x)-\overline{s}(x)] \neq 0$ \cite{Zeller:2002du, 
Mason:2004yf, Kretzer:2003wy}, and isospin violating PDFs \cite{Zeller:2002du, Kretzer:2003wy} 
may both be viable candidates for explaining the anomaly. 
Also the PDFs of bound nucleons, the nuclear parton distributions (nPDFs), 
have been considered as a source for the NuTeV anomaly \cite{Kovalenko:2002xe}; 
for a review, see \cite{McFarland:2003jw}. In particular the nuclear 
effects in the valence quark distributions $u_V$ and $d_V$ individually, 
have been thoroughly investigated in a global DGLAP analysis where the 
deep inelastic $lA$ scattering (DIS) cross sections, the Drell-Yan (DY) 
dilepton cross sections in $pA$ collisions and sum rules are used as constraints
\cite{Kumano:2002ra,Hirai:2004zu,Hirai:2004ba,Hirai:2006ww}. 
Typically, however, the effects of the nPDFs, especially the  
differences of nuclear modifications in $u_V$ relative to those in $d_V$,  
have been found to be too small to account for the NuTeV anomaly. 

In this paper, we consider the NuTeV anomaly and the nPDFs from a new angle, 
taking the NuTeV result as an additional data constraint for disentangling 
the nuclear effects of $u_V$ and $d_V$ distributions in nucleons 
(protons) of iron. First, adopting the EKS98 global analysis of nPDFs \cite{EKS98} 
as a baseline, we examine in a rough but transparent way, how large nuclear 
effects relative to the EKS98 modification of $u_V+d_V$ in a proton of iron 
would be needed to fully account for the NuTeV anomaly. 
Nuclear modifications for $u_V$ and $d_V$ deviating from the 
EKS98 at most on a $\sim 30$~\% level  (at the average NuTeV scale $Q^2=20.5$~GeV$^2$) 
turn out to be sufficient. 

Second, armed with the nuclear modifications for 
$u_V$ and $d_V$ individually, we recompute the DIS and DY cross sections, 
the data on which constrain the EKS98 modifications. 
Relative to the cross sections computed with EKS98, we find less than 
one percent changes for iron. This insensitivity shows that there 
can be fairly large differences between the nuclear effects in $u_V$ and $d_V$
which the standard global nPDF analyses \cite{EKS98,Kumano:2002ra,Hirai:2004ba}
cannot disentangle. It also
demonstrates the possible role of the NuTeV result as a new, orthogonal, constraint 
in pinning down the individual nuclear effects in the $u_V$ and $d_V$ distributions 
of bound nucleons. Not forgetting the other possible PDF-based explanations of the 
NuTeV anomaly, we thus suggest that a large part, if not all, of the NuTeV anomaly 
can be explained by introducing mutually different nuclear modifications for valence 
$u$ and $d$ quark distributions of protons in iron. 

Third, as a consequence and a test of our hypothesis, we apply the obtained 
nuclear valence quark modifications to the neutrino cross sections 
measured in the NOMAD experiment at CERN \cite{NOMAD}. 
We predict that the nonidentical nuclear effects in $u_V$ and $d_V$ 
induce a change in the ratio of neutral-to-charged current cross sections 
which, if identical nuclear effects were used, would correspond to an 
increase in $\sin^2\theta_W$ quite close to that obtained by NuTeV.

The rest of the paper is organized as follows: In Sec. 2, we define the 
neutrino cross sections and structure functions in terms of PDFs, as well 
as the Paschos-Wolfenstein (PW) ratio we study. In Sec. 3, we define the 
nPDFs and show how the nuclear $u_V$ and $d_V$ distributions enter the 
PW ratio. We also show analytically the direction to which the nuclear 
effects in $u_V$ and $d_V$ should deviate relative to the modification of 
$u_V+d_V$ from EKS98. The results we obtain for individual nuclear 
modifications of $u_V$ and $d_V$ are presented in Sec. 4 both for the 
case where the NuTeV anomaly is fully explained and for the case where 
a third of the anomaly is accounted for. Section 5 contains the discussion 
on the DIS and DY cross sections. The NOMAD prediction is made in Sec. 6, 
and a brief outlook is given in Sec. 7.

\section{Cross sections and Paschos-Wolfenstein ratio}
We define the framework through the following well-known charged current (CC) and neutral current 
(NC) DIS cross sections of neutrinos and antineutrinos off a nucleus. In particular,
we shall discuss the Paschos-Wolfenstein (PW) ratio \cite{Paschos:1972kj}. In terms of double 
differential cross sections, the PW ratio is  
\beq
R^-_A(x,Q^2) \equiv \frac{ d\sigma_A^{\nuNC}/dxdQ^2 - d\sigma_A^{\nubarNC}/dxdQ^2 }
{ d\sigma_A^{\nuCC}/dxdQ^2 - d\sigma_A^{\nubarCC}/dxdQ^2 }, 
\label{PWdiff}
\eeq
where $x$ and $Q^2$ are the standard Bjorken DIS variables, and $A$ denotes the target.
In studying the consequences of the NuTeV $\sin^2\theta_W$ anomaly quantitatively, 
we shall focus on the PW ratio of $x$-integrated cross sections at a fixed scale $Q^2$, 
\beq
R^-_A(x) \equiv \frac{ d\sigma_A^{\nuNC}/dQ^2 - d\sigma_A^{\nubarNC}/dQ^2}
{ d\sigma_A^{\nuCC}/dQ^2 - d\sigma_A^{\nubarCC}/dQ^2}.
\label{PWint}
\eeq

%%%%%%%%%%%%% NEUTRAL CURRENT XS's %%%%%%%%%%%%%%%%%%%%%%%%%%%%%%%%%%%%
The neutral current cross section of a deep inelastic scattering of a 
neutrino (antineutrino) off a proton is \cite{PDG}
\bea
\frac{d\sigma^{\nunubarNC}}{dxdQ^2}
&=&\frac{G_F^2}{\pi}\bigg(\frac{M_Z^2}{M_Z^2+Q^2}\bigg)^2\frac{1}{8x}
 \bigg[xy^2 F_1^{\nunubarNC}(x,Q^2) \nonumber \\
&+&  (1-y-\frac{mxy}{2E}) F_2^{\nunubarNC}(x,Q^2)\pm 
xy(1-\frac{y}{2})F_3^{\nunubarNC}(x, Q^2)\bigg],
\label{sigmaNC}
\eea
where $y$ is one of the standard Bjorken DIS variables and $E$ is the 
(anti)neutrino beam energy. For these DIS variables $Q^2=2xymE$, where $m$ is 
the mass of a nucleon. We shall use here $m=940$~MeV both for protons and for 
neutrons. The $Z$-boson mass 
is $M_Z = 91.1876$~GeV, and the Fermi coupling constant 
$G_F = 1.16637\cdot 10^{-5}$~GeV$^{-2}$, taken from \cite{PDG}.
The structure functions are in the leading order expressed in terms of the 
quark and antiquark number densities in the free proton as 
\bea
F_1^{\nuNC}(x,Q^2) &=& F_1^{\nubarNC}(x,Q^2) 
= \sum_{q=u,d,s,c,b} (R_q^2 + L_q^2) \big[ q(x,Q^2) + \overline q(x,Q^2) 
\big]  
\label{F1NC}
\\
F_2^{\nuNC}(x,Q^2) &=& F_2^{\nubarNC}(x,Q^2) = 2xF_1^{\nuNC}(x,Q^2) 
\label{F2NC}
\\
F_3^{\nuNC}(x,Q^2) &=& F_3^{\nubarNC}(x,Q^2) =2\sum_{q=u,d} (L_q^2-R_q^2) 
q_V(x,Q^2),  
\label{F3NC}
\eea
where, as usual, $L_q = \tau_3-2e_qx_W$ and $R_q=-2e_qx_W$. The weak mixing 
angle is 
$x_W\equiv \sin^2\theta_W$, the quark's electric charge is $e_q$ and its 
third component
of the weak-isospin is $\tau_3$. All quarks are treated as massless here, as 
is usually 
the case in the DGLAP-evolved PDFs we shall apply below.
The valence quark distribution is defined as $q_V(x,Q^2)\equiv 
q(x,Q^2)-\overline q(x,Q^2)$.

%%%%%%%%%% CHARGED CURRENT XS's %%%%%%%%%%%%%%%%%%%%%%

The charged current deep inelastic cross section of a neutrino (antineutrino) 
scattering off a proton can be expressed in a similar manner \cite{PDG},
\bea
\frac{d\sigma^{\nunubarCC}}{dxdQ^2}
&=&\frac{G_F^2}{\pi}\bigg(\frac{M_W^2}{M_W^2+Q^2}\bigg)^2\frac{1}{2x}
 \bigg[xy^2 F_1^{\nunubarCC}(x,Q^2) \nonumber \\
&+&  (1-y-\frac{mxy}{2E}) F_2^{\nunubarCC}(x,Q^2)\pm 
xy(1-\frac{y}{2})F_3^{\nunubarCC}(x, Q^2)\bigg].
\label{sigmaCC}
\eea
For the $W$ boson mass we take $M_W = 80.425$~GeV  \cite{PDG}.
The charged-current structure functions are given by 

\bea
F_1^{\nuCC}(x,Q^2) &=& \sum_{qq'}
\big[ q(x_{q'},Q^2) + \overline{q'}(x_{q},Q^2) \big] |V_{\rm CKM}^{qq'}|^2, 
\label{F1nuCC}
\\
F_1^{\nubarCC}(x,Q^2) &=& \sum_{qq'} 
\big[ \overline{q}(x_{q'},Q^2) + q'(x_{q},Q^2) \big] |V_{\rm CKM}^{qq'}|^2, 
\label{F1nubarCC}
\\
F_2^{\nuCC}(x,Q^2) &=& 2 \sum_{qq'}
\big[ x_{q'}q(x_{q'},Q^2) + x_{q}\overline{q'}(x_{q},Q^2) \big] |V_{\rm 
CKM}^{qq'}|^2,
\label{F2nuCC}
\\
F_2^{\nubarCC}(x,Q^2) &=& 2 \sum_{qq'}
\big[ x_{q'}\overline{q}(x_{q'},Q^2) + x_{q}q'(x_{q},Q^2) \big] |V_{\rm 
CKM}^{qq'}|^2, 
\label{F2nubarCC}
\\
F_3^{\nuCC}(x,Q^2) &=& 2 \sum_{qq'}
\big[ q(x_{q'},Q^2) - \overline{q'}(x_{q},Q^2) \big] |V_{\rm CKM}^{qq'}|^2, 
\label{F3nuCC}
\\
F_3^{\nubarCC}(x,Q^2) &=& 2 \sum_{qq'}
\big[ q'(x_{q},Q^2) - \overline{q}(x_{q'},Q^2) \big] |V_{\rm CKM}^{qq'}|^2, 
\label{F3nubarCC}
\eea
where the quark flavour indices  run through $q=d,s,b$ and $q'=u,c$. 
The momentum fraction $x_q \equiv x(1+\frac{m_q^2}{Q^2})$ accounts for the 
mass of the produced heavy quark. We consider $u$ and $d$ quarks massless 
and take $m_c =1.2$~GeV and $m_b = 4.3$~GeV. 
We also set $q(x_{q},Q^2)=0$ when $x_{q}\geq 1$.
In the initial state, the quark masses are ignored here, as is the case in 
the scale evolution of the PDFs, too. The Cabibbo-Kobayashi-Maskawa mixing matrix 
elements are denoted by 
$V_{\rm CKM}^{qq'}$. In what follows, we shall use
$V_{\rm CKM}^{du}=0.97458$, 
$V_{\rm CKM}^{dc}=0.224$,
$V_{\rm CKM}^{su}=0.224$,
$V_{\rm CKM}^{sc}=0.9738$,
$V_{\rm CKM}^{bu}=0.003$,
$V_{\rm CKM}^{bc}=0.04$, 
taken from \cite{PDG}.

For forming the Paschos-Wolfenstein ratios in Eqs.~(\ref{PWdiff}) and 
(\ref{PWint}), we form the differences of the above neutrino and 
antineutrino cross sections, 
\bea
\frac{d\sigma^{\nuNC}}{dxdQ^2} - \frac{d\sigma^{\nubarNC}}{dxdQ^2}  
&=&\frac{G_F^2}{\pi}\bigg(\frac{M_Z^2}{M_Z^2+Q^2}\bigg)^2\frac{1}{8x}
xy(1-\frac{y}{2})2F_3^{\nunubarNC}(x, Q^2),
\label{NCdiff}
\\
\frac{d\sigma^{\nuCC}}{dxdQ^2} - \frac{d\sigma^{\nubarCC}}{dxdQ^2}
&=&\frac{G_F^2}{\pi}\bigg(\frac{M_W^2}{M_W^2+Q^2}\bigg)^2\frac{1}{2x}
 \bigg[xy^2 \Delta F_1^{\rm CC}(x,Q^2) \nonumber \\
&+&  (1-y-\frac{mxy}{2E}) \Delta F_2^{\rm CC}(x,Q^2)+ 
xy(1-\frac{y}{2})\Sigma F_3^{\rm CC}(x, Q^2)\bigg],
\label{CCdiff}
\eea
where we define
\bea
\nonumber
\Delta F_1^{\rm CC}(x,Q^2) &\equiv& F_1^{\nuCC}(x,Q^2)-F_1^{\nubarCC}(x,Q^2) 
\\
&=& \sum_{qq'}
\big[ q_V(x_{q'},Q^2) - q'_V(x_q,Q^2) \big] |V_{\rm CKM}^{qq'}|^2, 
\label{F1diff}\\
\nonumber
\Delta F_2^{\rm CC}(x,Q^2)&\equiv& F_2^{\nuCC}(x,Q^2)-F_2^{\nubarCC}(x,Q^2) 
\\
&=& 2 \sum_{qq'}
\big[ x_{q'}q_V(x_{q'},Q^2) - x_q q'_V(x_q,Q^2) \big] |V_{\rm CKM}^{qq'}|^2, 
\label{F2diff}
\\
\nonumber
\Sigma F_3^{\rm CC}(x, Q^2) &\equiv& F_3^{\nuCC}(x,Q^2)+F_3^{\nubarCC}(x,Q^2) 
\\
&=& 2 \sum_{qq'} 
\big[ q_V(x_{q'},Q^2) + q'_V(x_q,Q^2) \big] |V_{\rm CKM}^{qq'}|^2, 
\label{F3diff}
\eea
where again the quark flavour indices run through  $q=d,s,b$ and $q'=u,c$.

\section{nPDFs and $R^-_A$}
We define the quark PDFs $Q_A(x,Q^2)$ in a nucleus of mass number $A$ and 
proton number $Z$ as in the EKS98 global DGLAP analysis \cite{EKS98}. Denoting the number 
density distribution of the quark flavour $q$ in a bound proton by $q^{p/A}(x,Q^2)$, 
and  the corresponding parton distribution function in a bound neutron by 
$q^{n/A}(x,Q^2)$, the average number density of a quark $q$ in the nucleus is  
\beq
Q_A(x,Q^2) = Zq^{p/A}(x,Q^2) + (A-Z)q^{n/A}(x, Q^2),
\label{nPDFsQ}
\eeq
Since we shall consider isospin effects below, it is useful to separate the 
symmetric and antisymmetric parts in the PDFs in the bound proton and neutron 
as
\beq
Q_A(x,Q^2) = \frac{A}{2}\left[q^{p/A}(x,Q^2)+q^{n/A}(x,Q^2)\right] +  
\frac{\Delta_A}{2}\left[q^{n/A}(x,Q^2)-q^{p/A}(x,Q^2)\right], 
\label{nPDFs_iso}
\eeq
where the neutron excess is denoted as $\Delta_A \equiv A-2Z$. In particular, 
as in EKS98,
we define the nuclear PDFs to be those of the bound protons, 
\beq
q_A(x,Q^2)\equiv q^{p/A}(x,Q^2),
\label{nPDFsq} 
\eeq 
and define their nuclear modifications relative to the corresponding PDFs in 
the free proton, 
\beq
q_A(x,Q^2) \equiv R_q^A(x,Q^2){q(x,Q^2)}.
\label{f_A}
\eeq

Again as in the EKS98 analysis \cite{EKS98}, we assume that isospin symmetry 
between the bound protons and neutrons holds for arbitrary $A$, i.e. $u^{n/A}=d^{p/A}$, 
$d^{n/A}=u^{p/A}$, and similarly for $\bar u$ and $\bar d$.
For mirror symmetric nuclei (which of course includes isoscalars), this gives 
the standard isospin symmetric way of treating the PDFs in protons and neutrons.
Thus, in the EKS98-framework adopted here, the isospin effects in 
the average quark distributions $Q_A$ are always proportional to the neutron 
excess $\Delta_A$ and to the difference between the $u(\bar u)$ and $d(\bar d)$ 
quark(antiquark) distributions in a bound proton, while the possible difference 
between e.g. the $u$ quark distribution in a proton of a nucleus $_Z^NA$ and 
the $u$ quark distribution in a proton of its mirror nucleus $_N^ZA$ is neglected. 
In other words, the EKS98 nuclear effects in the bound proton PDFs depend only on $A$,
not on $Z$. The scope of the presently existing data does not allow 
for a more detailed study of these effects. In practice, however, since we 
are dealing only with close-to-isoscalar nuclei, such as iron $(A=56$, $Z=26$) here, the 
isospin-symmetry-related approximation made can be expected to be a very good 
one. 

In particular, below we shall discuss the average valence quark modification,
\beq
R_V^A(x,Q^2) \equiv \frac{u_V^A(x,Q^2)+d_V^A(x,Q^2)}{u_V(x,Q^2)+d_V(x,Q^2)},
\label{RV}
\eeq
in terms of individual modifications for the  the valence $u$ quarks and 
valence $d$ quarks,
\bea
R_{u_V}^A(x,Q^2) &\equiv& \frac{u_V^A(x,Q^2)}{u_V(x,Q^2)} 
\label{RuV},\\
R_{d_V}^A(x,Q^2) &\equiv& \frac{d_V^A(x,Q^2)}{d_V(x,Q^2)}.
\label{RdV}
\eea
Only two of these ratios are independent, so that e.g. 
\beq
R_{u_V}^A(x,Q^2)= R_V^A(x,Q^2)\frac{u_V(x,Q^2)+d_V(x,Q^2)}{u_V(x,Q^2)}
-R_{d_V}^A(x,Q^2)\frac{d_V(x,Q^2)}{u_V(x,Q^2)}.
\label{RuV2}
\eeq

Analogously to the global DGLAP fits of the free proton PDFs,
in the global DGLAP fits for the nPDFs \cite{EKS98,Kumano,deFlorian}, 
the modification ratios $R_i^A(x,Q_0^2)$, $i=g,q$ at an initial scale $Q_0^2$ 
are iteratively determined based on fits to the deep inelastic $lA$ 
scattering data and the Drell-Yan dilepton data from $pA$ collisions. 
Also conservation of baryon number, charge and momentum are accounted for. 
The amount and precision of the existing data do not, however, 
allow for a precise determination of the modifications $R_{u_V}^A$ and $R_{d_V}^A$ 
individually but for simplicity in \cite{EKS98,Kumano,deFlorian}
it is assumed that at the initial scale  
$R_{u_V}^A=R_{d_V}^A=R_V^A$. Once this assumption is made, the DGLAP scale 
evolution does not cause large deviations from it, and to a good approximation 
$R_{u_V}^A=R_{d_V}^A=R_V^A$ holds for scales $Q^2\ge Q_0^2$ as well, 
see EKS98 \cite{EKS98}.  The most recent efforts 
\cite{Kumano:2002ra,Hirai:2004zu,Hirai:2004ba,Hirai:2006ww}
to disentangle the nuclear effects in $u_V$ and $d_V$ individually based on DIS 
and DY data do not lend support to significant differences between 
$R_{u_V}^A$ and $R_{d_V}^A$, either.

Below, we shall study to what extent the NuTeV $\sin^2\theta_W$ anomaly could 
be explained by releasing this approximation, i.e. by considering mutually 
different $R_{u_V}^A$ and $R_{d_V}^A$ but keeping their relative magnitude 
such that the EKS98-value of average valence modification, $R_V^A$, is always 
reproduced, as in Eq.~(\ref{RuV2}). In particular, we shall explore an 
extreme case where the whole NuTeV anomaly is accounted for by changes 
in $R_{u_V}^A$ and $R_{d_V}^A$, and study the possible consequences for the 
global DGLAP fits. 

To demonstrate the nuclear effects in the PW ratio 
$R_A^-(x)$, and to see the systematics expected, we first consider the ratio 
$R^-_A(x, Q^2)$ in an approximation, where only $u$ and $d$ quarks are 
active, $V_{\rm CKM}^{du}=1$, virtualities are modest, $Q^2\ll M_Z^2,M_W^2$, 
and beam energy is so high that the term $mxy/(2E)$ in Eqs. (\ref{sigmaNC}) 
and (\ref{sigmaCC}) can be ignored. 
Using the nPDF definitions above, we obtain for the neutral current case
\beq
F_{3,A}^{\nuNC}(x,Q^2) = A(2-4x_W)\left[u_V^A(x,Q^2) + d_V^A(x,Q^2)\right] + 
\Delta_A\frac{4}{3}x_W\left[u_V^A(x,Q^2) - d_V^A(x,Q^2)\right]
\eeq
and for the charged current case
\bea
\Delta F_{1,A}^{\rm CC}(x,Q^2) &=& \Delta_A\left[u_V^A(x,Q^2) - 
d_V^A(x,Q^2)\right] = \frac{1}{2x}\Delta F_2^{\rm CC}(x,Q^2)\\
\Sigma F_{3,A}^{\rm CC}(x,Q^2)&=& 2A\left[u_V^A(x,Q^2) + d_V^A(x,Q^2)\right].
\eea

Substituting these into Eqs.~(\ref{NCdiff}) and (\ref{CCdiff}) and forming 
the PW ratio in Eq.~(\ref{PWdiff}), we arrive at 
\bea
R^-_A(x,Q^2) = (\frac{1}{2}-x_W)
\bigg(1+\frac{\Delta_A}{A}\frac{2x_W}{3-6x_W}\frac{u_V^A-d_V^A}{u_V^A+d_V^A} \bigg)
\bigg(1+\frac{\Delta_A}{A}\frac{1+(1-y)^2}{1-(1-y)^2}\frac{u_V^A-d_V^A}{u_V^A
+d_V^A}\bigg)^{-1},
\label{PWapprox}
\eea
where the arguments $x,Q^2$ are implicit in the nPDFs on the r.h.s.

For isoscalar nuclei, $\Delta_A=0$, one obtains the conventional result, 
$R^-_{A_{\rm iso}}=1/2-x_W$. For non-isoscalar nuclei, isospin corrections 
proportional to $\Delta_A/A$ arise. We also immediately see that 
in the PW ratio of differential cross sections, $R^-_A(x,Q^2)$, the effects 
of nuclear valence quark modifications cancel out if the modifications are the same 
both for the nuclear valence-$u$ and for valence-$d$ quark distributions.

We compute the PW ratio in two different ways: 
\begin{itemize}
\item[1.] For $R_A^-(x,Q^2;x_W^{\rm NuTeV}, R_{d_V}^A=R_{u_V}^A=R_V^A)$,
we use the NuTeV value $x_W^{\rm NuTeV}=0.2277$ and take the {\em same} 
modification $R_V^A$ from EKS98 for both $R_{u_V}^A$ and $R_{d_V}^A$.
\item[2.] For $R_A^-(x,Q^2;\langle x_W\rangle, R_{d_V}^A\neq R_{u_V}^A)$,
we apply the world average value $\langle x_W\rangle =0.2227$ and assume  
$R_{u_V}^A\neq R_{d_V}^A$.
\end{itemize}
Setting these two PW ratios equal should then give us an idea into which 
direction the nuclear modifications $R_{u_V}^A$ and $R_{d_V}^A$ should 
deviate from the average valence modification $R_V^A$ so that the change 
in the PW ratio caused by the larger NuTeV value of $x_W$ is accounted for. 

Since the $y$ values we shall consider below, are always larger than 0.1 and 
$\Delta_A/A$ is of the order 0.1 for iron and $(u_V^A-d_V^A)/(u_V^A+d_V^A)\ll 1$, 
we can expand the denominator in Eq.~(\ref{PWapprox}). This gives 
\beq
R^-_A(x,Q^2,x_W)\approx (\frac{1}{2}-x_W)\bigg\{
1 - \frac{\Delta_A}{A}h(y,x_W)
\frac{u_V^A-d_V^A}{u_V^A+d_V^A},
\bigg\}
\eeq
where $h(y,x_W)\equiv \frac{1+(1-y)^2}{1-(1-y)^2} - \frac{2x_W}{3-6x_W}$
is always positive and not sensitive to small changes of $x_W$, since the 
first term dominates. Now, we set
\beq
R^-_A(x,Q^2;x_W^{\rm NuTeV}, R_{d_V}^A=R_{u_V}^A=R_V^A)
=
R^-_A(x,Q^2;\langle x_W\rangle, R_{d_V}^A\neq R_{u_V}^A)
\label{R=R}
\eeq
which, since $x_W^{\rm NuTeV}>\langle x_W\rangle$, leads to 
\beq
\frac{u_V-d_V}{u_V+d_V} < \frac{u_V^A-d_V^A}{u_V^A+d_V^A},
\eeq
where the nuclear effects on the l.h.s. have cancelled each other.
Then, applying the definitions in Eqs.~(\ref{RV})-(\ref{RuV2}) 
to the nuclear PDFs indicates that  
\beq
R_{d_V}^A(x,Q^2) < R_V^A(x,Q^2) <  R_{u_V}^A(x,Q^2)
\eeq
in the $x,Q^2$ region that the NuTeV cross sections are sensitive to.

\section{Results}

Let us now turn to a more detailed quantitative study of the PW ratio 
$R_A^-(x)$ in Eq.~(\ref{PWint}). In the NuTeV experiment \cite{Zeller:2001hh}, 
the neutrino and antineutrino cross sections were measured over a range of 
beam energies, virtualities $Q^2$ and Bjorken-$x$. Also cuts in the calorimeter 
energy were done. We do not attempt to simulate the kinematics in the experimentally 
collected event sample here in detail. Instead, we simply fix the beam energy 
to $E=116$~GeV, close to the NuTeV's average neutrino and antineutrino beam 
energies and the virtuality to $Q^2=20.5~\rm{GeV}^2$, corresponding to the 
NuTeV's average virtuality and integrate over $x$ in forming the PW ratio 
in  Eq.~(\ref{PWint}). To mimic the NuTeV cuts on the calorimeter energy, 
we require the final state energy on the hadron-remnant side to be in the range 
$E_{\rm min}\le m+yE\le E_{\rm max}$, with $E_{\rm min}=20$~GeV and 
$E_{\rm max}=180$~GeV. This, together with the requirement $y=Q^2/(2mEx)\le 1$, 
fixes the integration limits for $x$ as 
\beq
0.094\approx\frac{Q^2}{2mE}=x_{\rm min} \le x \le x_{\rm 
max}=\frac{Q^2}{2m(E_{\min}-m)}\approx0.57.
\eeq
We emphasize, however, that exact kinematic limits are somewhat irrelevant 
from the point of view of this paper. We aim to study whether 
order-of-magnitude changes in the nuclear modifications would be needed in 
order to explain the NuTeV anomaly in a very conventional manner. We shall 
show that in fact surprisingly modest modifications, of the order 20-30~\%, 
are sufficient. In the following, all the results are for iron nucleus $A=56$ 
and $\Delta_A=4$.

\begin{figure}[tbh]
 \begin{center}
\vspace{0cm}
    \epsfysize 7.0cm \epsfbox{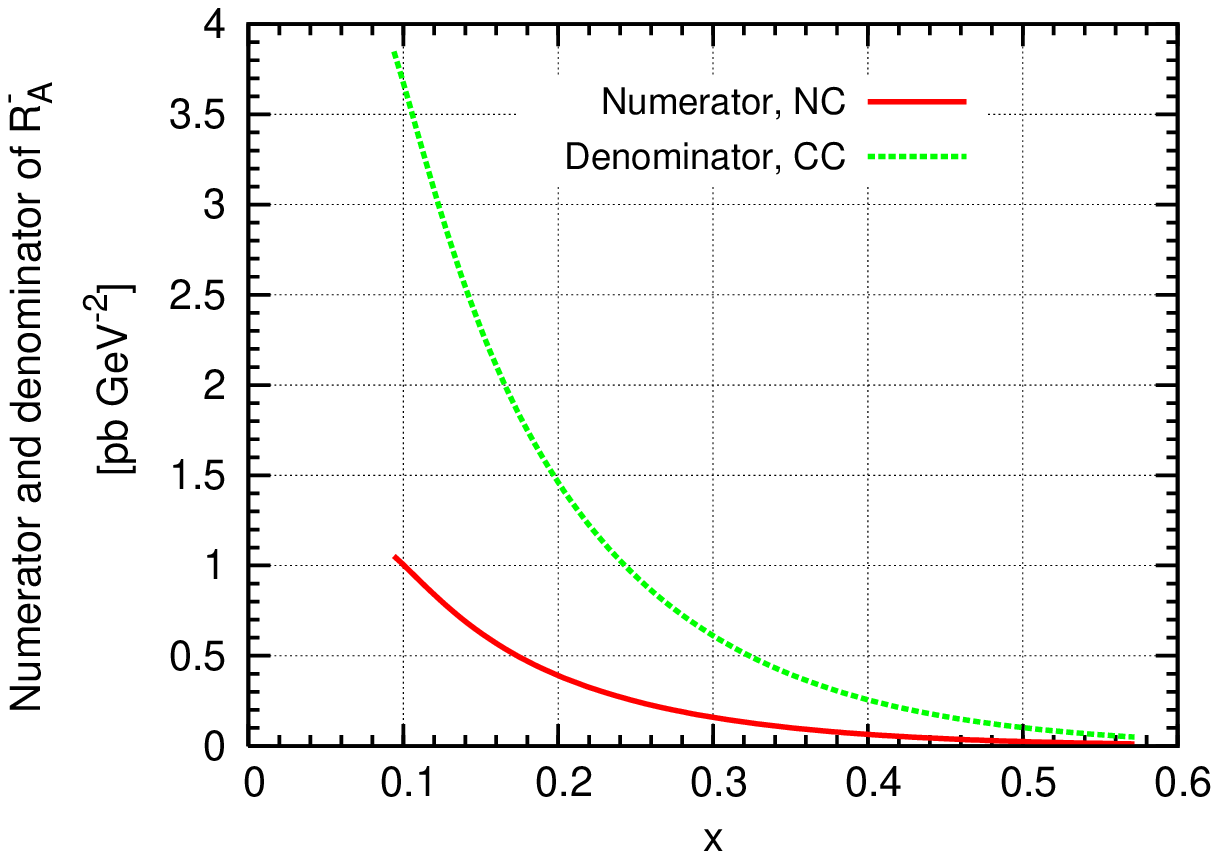}
 \vspace{-0mm}
\caption{\protect\small
The cross section differences 
$d\sigma^{\nuNC}_A/dxdQ^2-d\sigma^{\nubarNC}_A/dxdQ^2$
and $d\sigma^{\nuCC}_A/dxdQ^2-d\sigma^{\nubarCC}_A/dxdQ^2$  as a function of $x$
computed from Eqs.~(\ref{NCdiff}) and  (\ref{CCdiff}) at $Q^2=20.5$~GeV and 
$E=116$~GeV, with EKS98 \cite{EKS98} nuclear PDF effects for iron, $A=56$, 
$Z=26$, and using the world average value of mixing angle 
$\left\langle x_W\right\rangle=0.2227$.  
}
  \label{fig1:dsigmadx}
 \end{center}
\end{figure}

In Fig.~\ref{fig1:dsigmadx} we show the differences of the double 
differential neutrino and antineutrino cross sections from Eqs.~(\ref{NCdiff}) 
and (\ref{CCdiff}) as a function of $x$, computed with $\langle x_W\rangle$, 
and CTEQ6L \cite{CTEQ6L} PDFs with $R_{d_V}^A=R_{u_V}^A= R_V^A$ from EKS98 
\cite{EKS98}. From the result, it is obvious that the smallest-$x$ region 
dominates in the $x$-integrated cross sections. Thus, we should be looking 
for modifications of $R_{d_V}^A$ and $R_{u_V}^A$ in the region 
$x\gsim x_{\rm min}\sim 0.1$.

%%%%%%%%%%%%%%%%%%%%% FIGURE %%%%%%%%%%%%%%%%%%%%%%%%%%%%%%%%
%%%%%%%%%%%%%%%%%%%%%%%%%%%%%%%%%%%%%%%%%%%%%%%%%%%%%%%%%%%%%%

\begin{figure}[htb]
 \begin{center}
%\vspace{5cm}
    \epsfysize 7.0cm \epsfbox{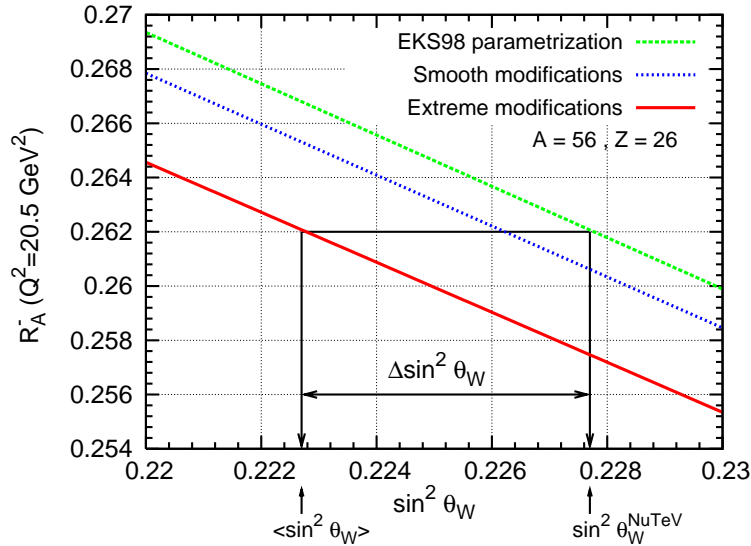}
% \vspace{-0mm}
\caption{\protect\small
The Paschos-Wolfenstein ratio as a function of the weak mixing angle, 
computed for iron, $A=56$, $Z=26$, by identifying the individual nuclear modifications 
$R_{u_V}^A$ and $R_{d_V}^A$ with the EKS98 $R_V^A$, 
with those in Fig.~\ref{fig2:Ruvall} ("extreme"), and with those
in Fig.~\ref{fig3:Ruv25} ("smooth"). The world average $\langle x_W\rangle$ and 
the NuteV result $x_W^{\rm NuTeV}$ are indicated.
}
  \label{fig1.5:proseduuri}
 \end{center}
\end{figure}

\begin{figure}[!]
 \begin{center}
%\vspace{-8mm}
    \epsfysize 7.0cm \epsfbox{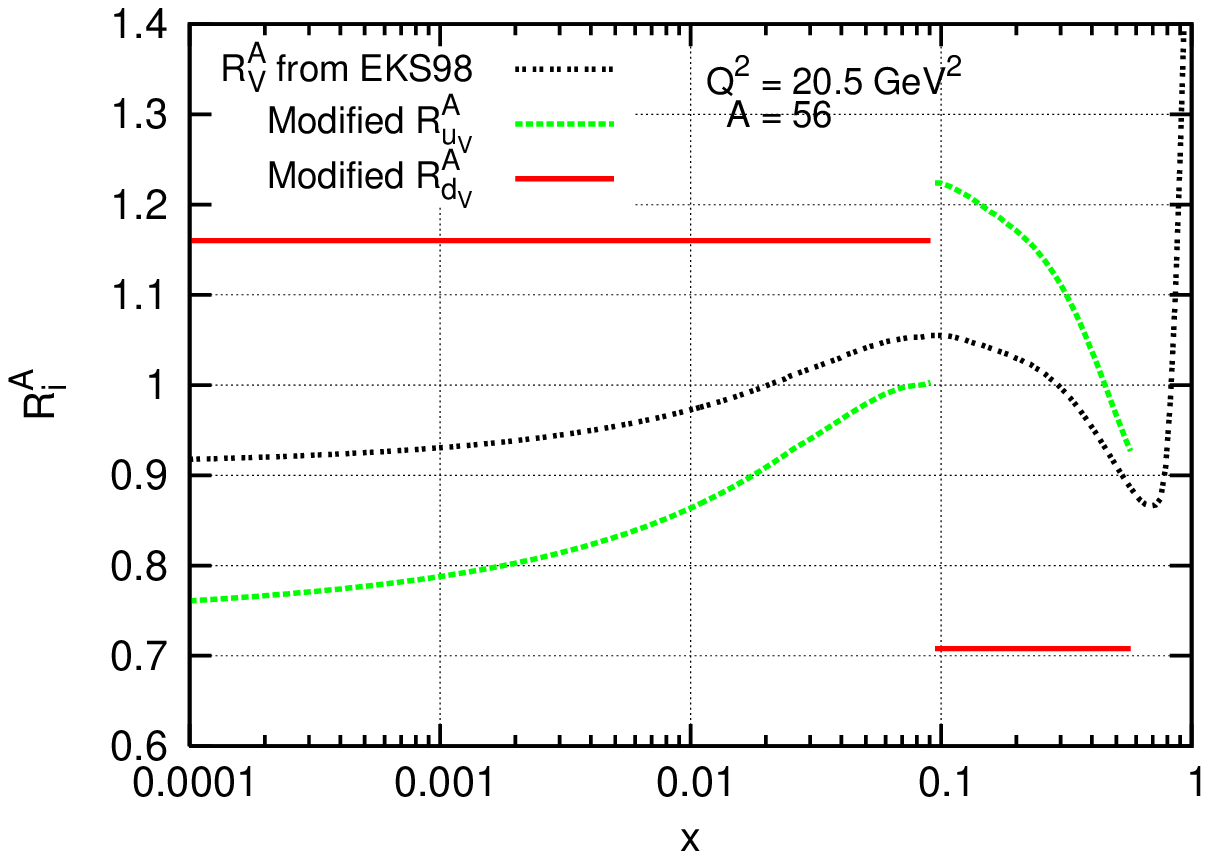}
% \vspace{-0mm}
%\label{fig2:Ruvall}
\caption{\protect\small
An order of magnitude estimate of the nuclear valence quark modifications
needed to alone account for the large value of $x_W$ reported by NuTeV.
The individual modifications $R_{u_V}^A$ and $R_{d_V}^A$ are shown for iron,
$A=56$, $Z=26$, as functions of $x$ at a fixed scale $Q^2=20.5$~GeV$^2$.
The average valence quark modification $R_V^A$ is from the 
EKS98-parametrization \cite{EKS98}.
Definitions of the ratios $R_i^A$ are given in Eqs. (\ref{RV})-(\ref{RdV}).
}
\label{fig2:Ruvall}
 \end{center}
%\end{figure}
%%%%%%%%%%%%%%%%%%%%%%%%%%%%%%%%%%%%%%%%%%%%%%%%%%%%%%%%%%%%%%
%%%%%%%%%%%%%%%%%%%%% FIGURE %%%%%%%%%%%%%%%%%%%%%%%%%%%%%%%%
%\begin{figure}[htb]
 \begin{center}
\vspace{1cm}
    \epsfysize 7.0cm \epsfbox{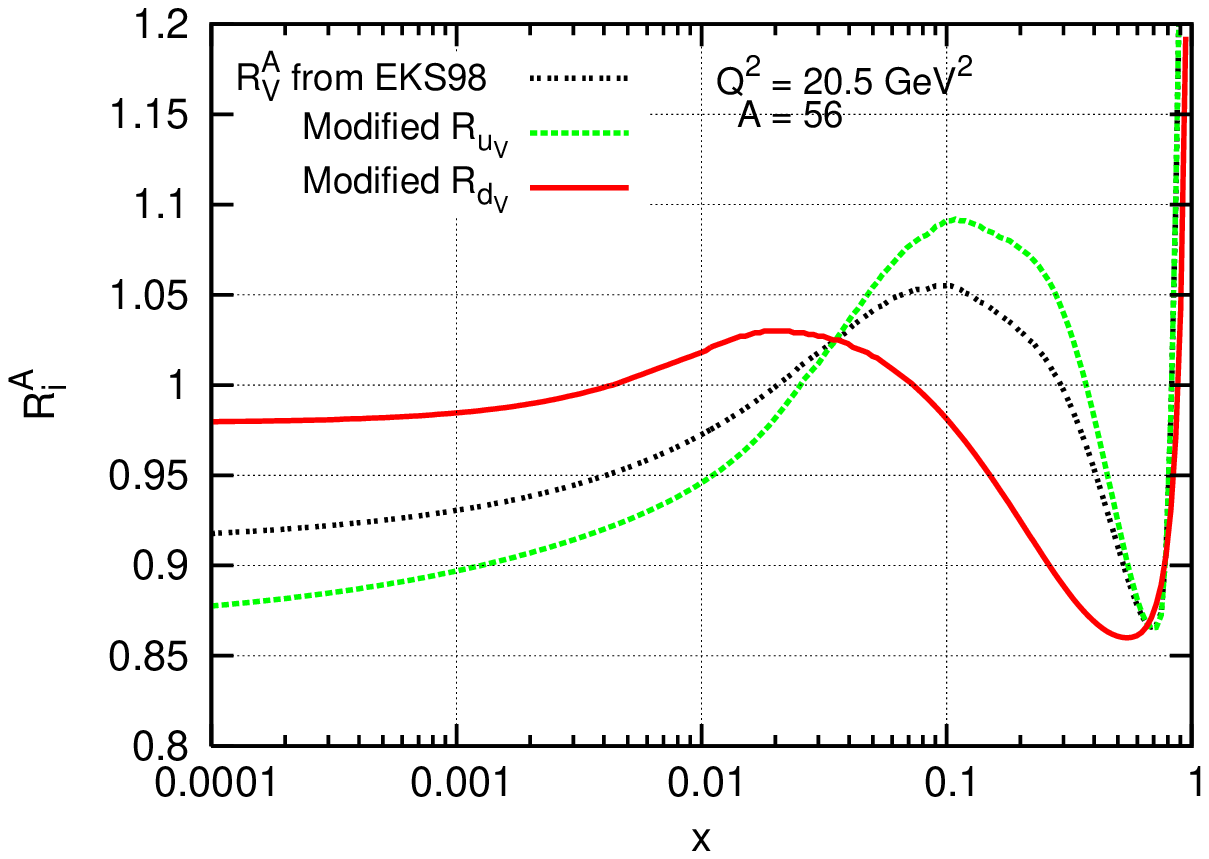}
% \vspace{-0mm}
\caption{\protect\small
As Fig.~\ref{fig2:Ruvall} but an example of a case where only one third of the 
difference $x_W^{\rm NuTeV}-\langle x_W\rangle$ would be explained.
}
  \label{fig3:Ruv25}
 \end{center}
\end{figure}

To get transparent order-of-magnitude estimates of the individual 
modifications $R_{d_V}^A$ and $R_{u_V}^A$ in light of the NuTeV 
$\sin^2\theta_W$ anomaly, we assume
the following very rough form for the valence-$d$ quark modification in iron
at $Q^2=20.5$~GeV$^2$ (see Fig.~\ref{fig2:Ruvall} ahead)
:
\beq
R_{d_V}^A(x,Q^2)= \left\{\begin{array}{ll}
C_1={\rm constant}, 			& \mbox{when $x<x_{\rm min}$}\\
C_2={\rm constant}, 			& \mbox{when $x_{\rm min}\le x\le x_{\rm 
max}$}\\
R_V^A(x,Q^2)\,{\rm from\, EKS98}	& \mbox{when $x_{\rm max}\le x \le 1$}
\label{RuvVextreme}
\end{array}
\right.
\eeq
For a fixed constant $C_2$, the constant $C_1$ is determined from charge 
conservation $\int_0^1 dx d_V^A(x,Q^2)=1$, after which the ratio $R_{u_V}^A$ 
is obtained from $R_V^A{(\scriptstyle \rm EKS98)}$ and $R_{d_V}^A$ according to 
Eq.~(\ref{RuV2}). Then, as the EKS98-result for $R_V^A$ conserves baryon 
number, charge conservation $\int_0^1 dx u_V^A(x,Q^2)=2$ is automatic. 
We proceed as outlined in the previous section. By using the $x$-integrated 
cross sections, we compute the PW ratios according to Eq.~(\ref{PWint}), on 
one hand for $x_W^{\rm NuTeV}$ and setting $R_{d_V}^A=R_{u_V}^A=R_V^A$, and 
on the other hand for $\langle x_W\rangle$
and setting $R_{d_V}^A\neq R_{u_V}^A$. The requirement (analogous 
to Eq.~(\ref{R=R}))
\beq
R_A^-(Q^2,x_W^{\rm NuTeV}, R_{d_V}^A=R_{u_V}^A=R_V^A)
=
R_A^-(Q^2,\langle x_W\rangle, R_{d_V}^A\neq R_{u_V}^A),
\label{explainNuTeV}
\eeq
now fixes the constant $C_2\approx 0.708$ in $R_{d_V}^A$, 
and consequently $C_1\approx 1.16$. 
Figure \ref{fig1.5:proseduuri} shows the PW ratio as a function of the weak 
mixing angle, computed with the mutually different nuclear modifications 
$R_{d_V}^A$ and $R_{u_V}^A$ (r.h.s. of Eq.~(\ref{explainNuTeV}), "extreme"), 
and with the EKS98 modification $R_V^A$ only 
(l.h.s. of Eq.~(\ref{explainNuTeV}), "EKS98").

%%%%%%%%%%%%%%%%%%%%% FIGURE %%%%%%%%%%%%%%%%%%%%%%%%%%%%%%%%
%%%%%%%%%%%%%%%%%%%%%%%%%%%%%%%%%%%%%%%%%%%%%%%%%%%%%%%%%%%%%%

%%%%%%%%%%%%%%%%%%%%% FIGURE %%%%%%%%%%%%%%%%%%%%%%%%%%%%%%%%
%\begin{figure}[htb]
%%%%%%%%%%%%%%%%%%%%%%%%%%%%%%%%%%%%%%%%%%%%%%%%%%%%%%%%%%%%%%

Fig.~\ref{fig2:Ruvall}, showing the extreme case where the mutually different
nuclear modifications to $u_V$ and $d_V$ explain the whole NuTeV anomaly, 
is the main result of this paper. Perhaps somewhat surprisingly, 
our result shows that even if we do not invoke any exotic new phenomena, 
such as isospin-breaking PDFs, locally non-zero $s-\bar s$ distributions, 
not to mention beyond-Standard-Model physics, to explain the NuTeV anomaly, 
fairly modest, $\sim 20-30$~\% changes in the nuclear valence quark modifications 
would seem to explain the large value of $x_W$ observed by NuTeV.

In addition to the extreme case considered above, we also consider, as an example, 
more modest changes with a continuous and smooth  parametrization for $R_{d_V}^A$ 
shown in Fig.~\ref{fig3:Ruv25}. 
Also here charge is conserved and the EKS98 ratio $R_{V}^A$ is reproduced as 
explained above. Proceeding again in the same manner as above, we find that 
these nuclear effects would correspond to $x_W=0.2243$, which would explain 
a third of the NuTeV anomaly $\Delta x_W = 0.005$.

\section{Effects on $R_{F_2}^A$ and $R_{\rm DY}^A$}

Next, we show that the  modifications $R_{u_V}^A$ and 
$R_{d_V}^A$ obtained above do not induce severe modifications into the 
quantities which are already rather well constrained in the global DGLAP 
analyses of nPDFs. As mentioned above, the data constraints are given by the 
$lA$ DIS cross sections and Drell-Yan dilepton cross sections in $pA$ 
collisions. 

\begin{figure}[!]
  \begin{center}
%\vspace{-8mm}
    \epsfysize 7.0cm \epsfbox{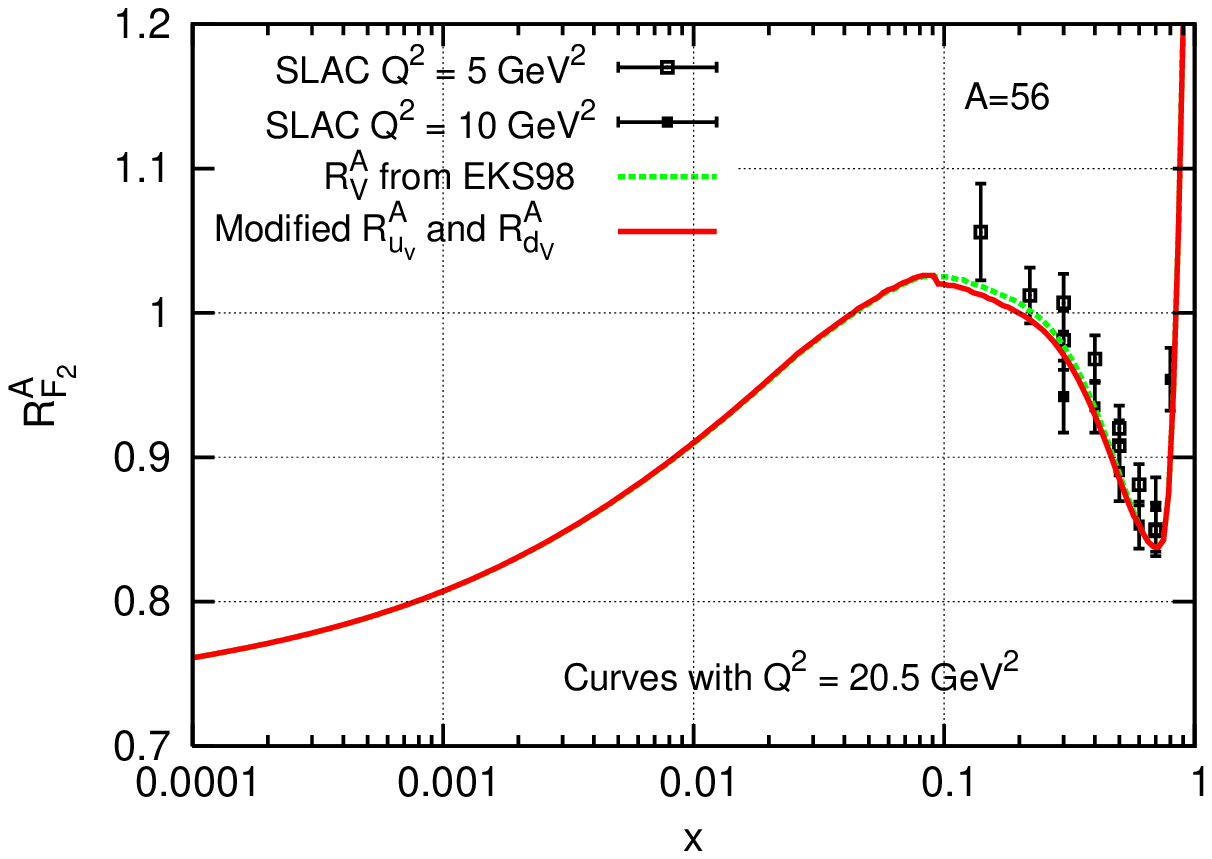}
% \vspace{-0mm}
\caption{\protect\small
The nuclear modification ratio $R_{F_2}^A$ for non-isospin-symmetrized iron, 
$A=56$ and $Z=26$, as a function of $x$ at a fixed scale $Q^2=20.5$~GeV$^2$. 
The dashed curve is computed with the EKS98 modifications and the solid one 
is the result when the extreme modifications of Fig.~\ref{fig2:Ruvall} are 
applied for $R_{u_V}^A$ and $R_{d_V}^A$. The SLAC data shown have been isospin 
symmetrized \cite{Gomez:1993ri}.}
  \label{fig4:RF2}
 \end{center}
%\end{figure}
%%%%%%%%%%%%%%%%%%%%%%%%%%%%%%%%%%%%%%%%%%%%%%%%%%%%%%%%%%%%%%
%%%%%%%%%%%%%%%%%%%%% FIGURE %%%%%%%%%%%%%%%%%%%%%%%%%%%%%%%%
%\begin{figure}[htb]
 \begin{center}
\vspace{1cm}
    \epsfysize 7.0cm \epsfbox{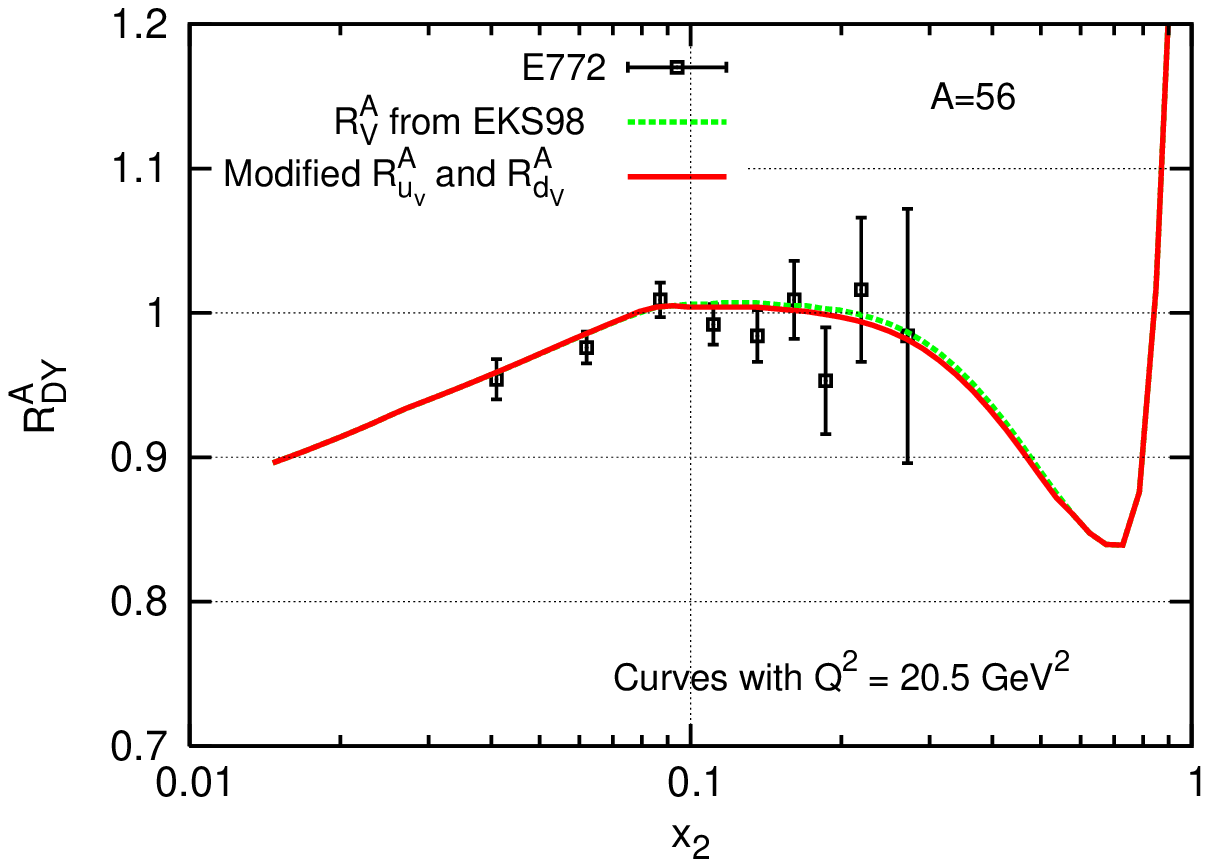}
% \vspace{-0mm}
\caption{\protect\small
The nuclear modification ratio $R_{\rm DY}^A(x_2,Q^2)$ for 
non-isospin-symmetrized iron, $A=56$ and $Z=26$,
as a function of the target-side momentum fraction $x_2$ at a fixed scale 
$Q^2=20.5$~GeV$^2$. The dashed curve is computed with the EKS98 modifications 
and the solid one is the result when the extreme modifications of 
Fig.~\ref{fig2:Ruvall} are used for $R_{u_V}^A$ and $R_{d_V}^A$. 
The data shown are from E772 \cite{Alde:1990im}.}
  \label{fig5:RDY}
 \end{center}
\end{figure}

In DIS of leptons off the free proton, the leading order structure functions 
are 
\beq
F_2^{lp}(x,Q^2) = \sum_q e_q^2x\bigg[ q(x,Q^2) + \bar q(x,Q^2)  \bigg] = 
2xF_1^{lp}(x,Q^2).
\eeq
The nuclear modifications of $F_2^{lA}$ are specified relative to deuterium,
\beq
R_{F_2}^A(x,Q^2) 
\equiv 
\frac{\frac{1}{A}d\sigma^{lA}/dxdQ^2}{\frac{1}{2}d\sigma^{lD}/dxdQ^2} 
= 
\frac{\frac{1}{A}F_2^{lA}(x,Q^2)}{\frac{1}{2}F_2^{lD}(x,Q^2)}, 
\label{RF2}
\eeq
Then, as discussed in \cite{EKS98}, and using the definitions of nPDFs given 
here, the nuclear modifications of $F_2^{lA}$ relative to deuterium become 
(suppressing the arguments $x,Q^2$)
\beq
R_{F_2}^A = \frac{5(u_A+d_A + \bar u_A + \bar d_A) + 4(s_A+\bar s_A) + \dots 
+ \frac{3\Delta_A}{A}(d_A+\bar d_A -u_A - \bar u_A)}
{5(u+d + \bar u + \bar d) + 4(s+\bar s) + \dots},
\label{RF2EKS98}
\eeq
where the dots denote the heavy quark contributions and the small nuclear 
effects in deuterium have been neglected. From above, we see that since 
$u_A+d_A=u_V^A+d_V^A + \bar u_A + \bar d_A 
= R_V^A(u_V+d_V) + \bar u_A + \bar d_A $, any change in 
$R_{u_V}^A$ and $R_{d_V}^A$ that leaves
the valence modification $R_V^A$ unchanged, will not at all change the 
isoscalar part. We emphasize that the DIS $lA$ data has usually been approximately 
"isospin-symmetrized" by the experimental collaborations. E.g. the EKS98 analysis 
is using such isospin-symmetrized DIS data (NMC, SLAC mainly) and therefore in 
constraining the fits, one is only discussing isospin symmetric nuclei for DIS. 
We show in Fig.~\ref{fig4:RF2} the effects that our 
extreme modifications of $R_{d_V}^A$ and $R_{u_V}^A$ in Fig.~\ref{fig2:Ruvall}
would induce into the ratio $R_{F_2}^A$ of non-isoscalar iron 
at $Q^2=20.5$~GeV$^2$. Since the changes due to $R_{d_V}^A\neq R_{u_V}^A$
only take place in the subleading correction term proportional to 
$\Delta_A/A$, the effects remain very small, below one per cent level for 
iron. 
Given the precision of the existing data, such small changes can be easily 
accommodated in the global DGLAP fits. 

%%%%%%%%%%%%%%%%%%%%% FIGURE %%%%%%%%%%%%%%%%%%%%%%%%%%%%%%%%
%\begin{figure}[htb]
%%%%%%%%%%%%%%%%%%%%%%%%%%%%%%%%%%%%%%%%%%%%%%%%%%%%%%%%%%%%%%

Finally, we perform a similar check for the Drell-Yan nuclear modification 
ratio $R_{\rm DY}^A(x_2,Q^2)$ (see \cite{EKS98}), which can be written in the 
LO as follows
\bea
\nonumber
R_{\rm DY}^A(x_2,Q^2)\equiv\frac{\frac{1}{A}d\sigma^{pA}/dx_2dQ^2}
{\frac{1}{2}d\sigma^{pD}/dx_2dQ^2}
\hspace{10cm}\\
= \{
4[u_1(\bar u_2^A+\bar d_2^A) + \bar u_1(u_2^A+d_2^A)]
+[d_1(\bar d_2^A+\bar u_2^A)+ \bar d_1(u_2^A+d_2^A)] + 4s_1s_2^A + \dots
\}/N_{\rm DY}\hspace{0.5cm}\nonumber\\
+ \frac{\Delta_A}{A}\{
4[u_1(\bar d_2^A-\bar u_2^A) + \bar u_1(d_2^A-u_2^A)]
+[d_1(\bar u_2^A-\bar d_2^A)+ \bar d_1(u_2^A-d_2^A)] 
\}/N_{\rm DY},\hspace{0.5cm}
\eea
where $N_{\rm DY}= 4[u_1(\bar u_2+\bar d_2) + \bar u_1(u_2+d_2)]
+[d_1(\bar d_2+\bar u_2)+ \bar d_1(u_2+d_2)] +4s_1s_2 + \dots$ and 
$Q^2$ is the invariant mass of the lepton pair, and where we have used 
the notation
$q_j^{(A)}\equiv q_{(A)}(x_j,Q^2)$ with  $j=1,2$, where $x_2(x_1)$ is the 
fractional momentum of the colliding parton from the target (projectile).
Again the quark combination $u_A+d_A$ appears in the leading, isoscalar, 
term. For isoscalars, any $R_V^A$-conserving changes made for $R_{u_V}^A$ and 
$R_{d_V}^A$ will not affect the ratio $R_{\rm DY}^A(x_2,Q^2)$ at all. Contrary 
to the DIS case, however, the DY data available are not isospin-symmetrized 
but show also some isospin effects. The modifications of 
$R_{u_V}^A$ and $R_{d_V}^A$ will be transmitted to $R_{\rm DY}^A$ only 
through the correction term proportional to $\Delta_A/A$ and hence even 
the extreme modifications of  Eq.~(\ref{RuvVextreme}) and 
Fig.~\ref{fig2:Ruvall} cause only a small deviation from the result 
computed with the EKS98 nuclear modifications. This is shown 
in Fig.~\ref{fig5:RDY}, where we compute the ratio $R_{\rm DY}^A(x_2,Q^2)$ as 
a function of $x_2$ at a fixed scale $Q^2=20.5$~GeV$^2$ 
on one hand with the extreme modifications for $R_{u_V}^A$ and $R_{d_V}^A$ and 
on the other with the EKS98 modifications only. 
Again the difference to the EKS98-based result is less than a percent. We thus 
conclude that the DIS and DY cross sections used in the global 
DGLAP fits are essentially insensitive to the decomposition of $R_V^A$ into 
$R_{u_V}^A$ and $R_{d_V}^A$, as long as the average valence quark 
modification remains unchanged. Thus, the interpretation of the NuTeV 
$\sin^2\theta_W$ anomaly in terms of conventional physics, mutually different 
$R_{u_V}^A$ and $R_{d_V}^A$, does not lead into a contradiction with the existing 
global DGLAP fits for the nPDFs.

\section{Effects on $R_A^{\nu}$ in NOMAD}
\begin{figure}[htb]
 \begin{center}
%\vspace{5cm}
    \epsfysize 7.0cm \epsfbox{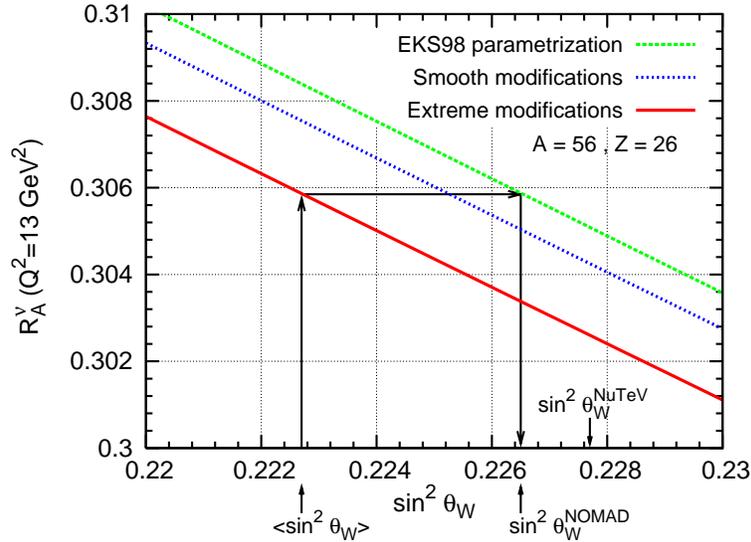}
% \vspace{-0mm}
\caption{\protect\small
The ratio $R^{\nu}_A$ for iron, $A=56, Z=26$,
as a function of $\sin^2\theta_W$, 
computed in the $x$-range of NOMAD at $Q^2=13$~GeV$^2$
with the EKS98 nuclear modifications ("EKS98"), 
with extreme individual nuclear modifications 
$R_{u_V}^A$ and $R_{d_V}^A$ in Fig.~\ref{fig2:Ruvall}
together with the EKS98 for other parton flavours ("extreme"), 
and with the smooth modifications in Fig.~\ref{fig3:Ruv25}
together with the EKS98 ("smooth").
The world average $\langle x_W\rangle$ and 
the expected increase of $x_W$ in the NOMAD case,
cf. Eq.~(\ref{predictNOMAD}), are indicated.
}
  \label{fig:NOMAD}
 \end{center}
\end{figure}

As the final point in this paper, we make the following prediction 
for the NOMAD neutrino experiment at CERN \cite{NOMAD}. In this experiment one 
can measure the ratio of charged current and neutral current neutrino-iron 
cross sections, 
\beq
R^{\nu}\equiv\frac{\sigma_A^{\nu,{\rm NC}}}{\sigma_A^{\nu,{\rm CC}}}.
\eeq 
The differential cross sections are given in Eqs.~(\ref{sigmaNC}) 
and (\ref{sigmaCC}). 
With the parameters $\langle Q^2\rangle=13$~GeV$^2$ and 
$\langle E_{\nu}\rangle=45.4$~GeV, the typical NOMAD $x$-range becomes
$x\gsim 0.15$. This is close to that of NuTeV, and the valence quarks 
dominate the $x$-integrated cross sections. As shown in Fig.~\ref{fig:NOMAD}, 
we compute the $x$-integrated cross sections at a fixed scale $Q^2=13$~GeV$^2$ 
and form the ratio $R^\nu_A$ again in two different ways: 
on one hand we apply the EKS98 nuclear effects, where $R_{u_V}^A=R_{d_V}^A=R_V^A$, 
and on the other we apply the individual modifications $R_{u_V}^A \neq R_{d_V}^A$  
from Figs.~\ref{fig2:Ruvall} and \ref{fig3:Ruv25} (and EKS98 for 
the parton flavours other than valence quarks). The figure demonstrates that 
the situation with $R^{\nu}_A$ in NOMAD should be very similar to that 
with the PW ratio in NuTeV:
if identical nuclear corrections are used for $u_V$ and $d_V$, the value of 
$x_W$ extracted from the NOMAD data will be larger than the world average
$\langle x_W \rangle$. If the whole NuTeV anomaly is explained by 
the individual, mutually different, nuclear effects in $u_V$ and $d_V$ distributions, then
solving $x_W^{\rm NOMAD}$ from 
\beq
R_A^{\nu}(Q^2,x_W^{\rm NOMAD}, R^{A,{\rm EKS98}}_i)
=
R_A^{\nu}(Q^2,\langle x_W\rangle, R_{i\neq u_V,d_V}^A
=R_i^{A,{\rm EKS98}}, R_{d_V}^A\neq R_{u_V}^A),
\label{predictNOMAD}
\eeq
where the individual modifications $R_{d_V}^A$ and $R_{u_V}^A$ are the extreme ones from 
Fig.~\ref{fig2:Ruvall}, gives $x_W^{\rm NOMAD}\approx 0.2265$, 
i.e. $\Delta x_W^{\rm NOMAD}=x_W^{\rm NOMAD}-\langle x_W\rangle \approx 0.004$, 
close to the result obtained by NuTeV.
With the smooth modifications in Fig.~\ref{fig3:Ruv25}, the increase in $x_W$ is again 
close to that obtained in Fig.~\ref{fig1.5:proseduuri}. 

%%%%%%%%%%%%%%%%%%%%% FIGURE %%%%%%%%%%%%%%%%%%%%%%%%%%%%%%%%
%%%%%%%%%%%%%%%%%%%%%%%%%%%%%%%%%%%%%%%%%%%%%%%%%%%%%%%%%%%%%%

\section{Conclusions}
For the nuclear valence quark PDFs one usually makes the approximation 
$u_V^A/u_V \approx d_V^A/d_V\approx (u_V^A+d_V^A)/(u_V+d_V)\equiv R_V^A$.
In this case, the effects from the nuclear valence quark modification $R_V^A$ 
are cancelled out in the Paschos-Wolfenstein ratio $R^-(x,Q^2)$. 
In this paper, we are suggesting that taking 
$R_{u_V}^A\equiv u_V^A/u_V\neq d_V^A/d_V\equiv R_{d_V}^A$ and 
by allowing nuclear effects of the order of 30 \%, 
leads to the modifications of the Paschos-Wolfenstein ratio $R^-(Q^2)$ 
which are as large as would be induced by increasing the weak mixing angle 
$x_W$ to the large value reported by NuTeV. We have also checked that as long 
as the valence modification $R_V^A$ is conserved in the decomposition
into $R_{u_V}^A$ and $ R_{d_V}^A$, the computed ratios of the nuclear DIS and 
DY  cross sections over deuterium, used as data constraints in the 
global DGLAP analyses of nPDFs, change by less than a percent in the case of 
iron. 
On one hand, this insensitivity shows that the explanation of the NuteV anomaly in terms 
of nPDFs is well possible without running into contradiction with the global nPDF analyses.
On the other hand, it shows, in agreement with 
\cite{EKS98,Kumano:2002ra,Hirai:2004ba}, that in practice
it is not possible to pin down the individual nuclear effects $R_{u_V}^A$ and $R_{d_V}^A$
based on finite-precision nuclear DIS and DY data. 
The main point of the present paper thus is that most, if not all, of the 
$\sin^2\theta_W$ NuTeV anomaly could be explained by conventional physics
and that the NuTeV result can play a key role in disentangling these effects 
for $u_V$ and $d_V$. Finally, as a consequence, we predict that if the whole 
NuTeV anomaly can be accounted for by the nuclear corrections as we suggest here,
and if identical nuclear effects for $u_V$ and $d_V$ are used in the data analysis,  
the value of $\sin^2\theta_W$ to be determined in the NOMAD neutrino experiment at CERN 
should be quite close to that obtained by NuTeV.

We wish to emphasize that in the global nPDF analyses the shapes of the individual nuclear 
corrections to the PDFs are determined based on the data constraints (through DGLAP evolution) 
and sum rules. The origin of these corrections, and PDFs in general, is nonperturbative.
If the NuTeV anomaly indeed is due to the nPDF effects as we suggest above, 
it would be very interesting to understand the mechanism which makes the nuclear effects for 
$u_V$ and $d_V$ distributions so different. We do not have an answer to this question for the 
moment.

In this paper, we do not make an attempt to simulate the NuTeV kinematics in detail but 
our goal is to show that interpretation of the NuTeV anomaly in terms of nPDF effects is
indeed possible and to see the order of magnitude for the effects needed. The more detailed 
kinematics, along with scale evolution details of the valence quark effects, we shall consider 
elsewhere. 

\vspace{1cm}
\noindent{\bf Acknowledgements.}
We thank P. Castorina,  V. Ruuskanen and K. Tuominen for discussions.
Financial support from the Academy of Finland, the Project No. 206024,
is gratefully acknowledged.

\end{document}